\documentclass{aa}
\usepackage{graphicx}
\usepackage{txfonts}
\usepackage{natbib}
\bibpunct{(}{)}{;}{a}{}{,}    
\newcommand{\beq}{\begin{equation}}
\newcommand{\eeq}{\end{equation}}
\newcommand{\bea}{\begin{eqnarray}}
\newcommand{\eea}{\end{eqnarray}}

\newcommand{\subscr}[1]{_\mathrm{#1}}

\newcommand{\url}[1]{{\tt #1}}

\newcommand{\doverdt}[1]{\frac{\partial #1}{\partial t}}
\def\gapp{\lower 3pt\hbox{${\buildrel > \over \sim}$}\ }
\def\lapp{\lower 3pt\hbox{${\buildrel < \over \sim}$}\ }

\begin{document}
\title{Migration of Protoplanets in Radiative Disks}
\author{
Wilhelm Kley  
\and
Aur\'elien Crida 
}
\offprints{W. Kley,\\ \email{wilhelm.kley@uni-tuebingen.de}}
\institute{
     Institut f\"ur Astronomie \& Astrophysik, 
     Universit\"at T\"ubingen,
     Auf der Morgenstelle 10, D-72076 T\"ubingen, Germany
}
\date{June 2008}
\abstract
{In isothermal disks the migration of protoplanets is directed inward. For small planetary
masses the standard type~I migration rates are so fast that this may
result in an unrealistic loss of planets into the stars.}
{We investigate the planet-disk interaction in non-isothermal disks and analyze the magnitude
and direction of migration for an extended range of planet masses.
}
{
We have performed detailed two-dimensional numerical simulations of embedded
planets including heating/cooling effects as well as radiative diffusion for
realistic opacities.
}
{In radiative disks, small planets with $M_{\rm planet} < 50 M_{\rm Earth}$
do migrate outward with a rate
comparable to absolute magnitude of standard type~I migration. 
For larger masses the migration is inward and approaches the isothermal, type~II migration rate.
}
{Our findings are particularly important for the first growth phase of planets and ease the problem
of too rapid inward type-I migration. 
}
\keywords{accretion disks -- planet formation -- hydrodynamics}
\maketitle
\markboth
{Kley \& Crida: Planetary migration in radiative disks}
{Kley \& Crida: Planetary migration in radiative disks}

\section{Introduction}
\label{sec:introduction}
Planets form in disks surrounding young stars. The growing
protoplanets undergo an embedded phase where the
gravitational interaction with the ambient gaseous disk results
in a change of its orbital elements.
For protoplanets with masses below about 30 Earth masses the
disk is not disturbed too strongly and the interaction can be treated in 
the linear approximation. Calculations of the total disk torques acting on the
planet lead generally for these small masses to a reduction of the semi-major
axis, i.e. to an inward migration 
\citep{1979ApJ...233..857G, 1997Icar..126..261W, 2002ApJ...565.1257T, 2004ApJ...602..388T}.
It soon turned out that the inward drift
of this type-I migration is very fast and the planets might be lost before they
can grow to larger objects \citep{1993Icar..102..150K}. 
This problem has become more visible after comparing
population synthesis models with the characteristics of observed planetary systems
\citep{2004A&A...417L..25A, 2008ApJ...673..487I}.
To avoid this rapid phase of inward migration, alternative scenarios have been
sought. In a turbulent disk, migration occurs stochastically with inward and
outward phases which slows down the migration \citep{2005A&A...443.1067N}.
Departures from the linear regime at around 10-20 $M_{\rm Earth}$ can also lead to reduced  
inward migration \citep{2006ApJ...652..730M}.
However, both processes are not sufficient to solve the problem.
The planet trap scenario to halt planetary migration \citep{2006ApJ...642..478M}
requires a positive density gradient which may not be given in general. 

To simplify the calculations, nearly all of the analytical and numerical studies
devoted to study the planet-disk interaction process have focussed on isothermal
disks, where the temperature is a given function of the position in the disk.
Early work on non-isothermal disks focussed on high mass embedded planets 
and did not notice a strong effect on migration \citep{2003ApJ...599..548D,
2006A&A...445..747K}.
Using a fully three-dimensional radiative calculations of an embedded small mass planet,
\citet{2006A&A...459L..17P} have shown in a very important work
that migration can be significantly slowed down or even reversed when thermal
effects are included.
Subsequent analysis indicate that this behaviour is related to a radial entropy gradient
in the flow \citep{2008ApJ...672.1054B, 2008A&A...478..245P}.
Recently, \citet{2008arXiv0804.4547P} have shown that for small planet
masses, the combination of radiative and viscous diffusion may allow
for long-term unsaturated positive torques and possible outward migration.

In this letter we investigate this possibility in more detail for a whole range of
planetary masses, which will allow us to estimate its effect on the long term
evolution of the planet.
For that purpose we perform two-dimensional numerical hydrodynamical simulations
of embedded planets in radiative disks. A method to treat the three-dimensional
radiative transfer approximately in these 2D simulations will be outlined in the
next section. Our results on the migration rate for various masses (in Section 3) indicate
that for masses smaller than about 50~$M_{\rm Earth}$ the torques remain unsaturated in the
long run and migration is indeed directed outward, while larger planets drift inward.
The consequence for the migration process and the overall evolution of planets
in disks is discussed.

\section{Physical modelling}
\label{sec:model}
The protoplanetary disk is treated as a two-dimensional,
non-self-gravitating gas that can be described by the Navier-Stokes
equations.  The embedded planet is modelled as a point mass that
orbits the central star on fixed, circular orbit. For the planetary
potential we use a smoothing length of $\epsilon = 0.6 H$ where $H$ is
the vertical scale height of the disk.  To calculate the gravitational
torques acting on the planet we apply a tapering function to exclude
the inner parts of the Hill sphere of the planet, where the transition
lies at $0.8$ Hill radii \citep[see][ Fig.~2]{2008A&A...483..325C}.
\subsection{Energy Equation}
In the present setup we utilise fully radiative models with an
improved thermodynamic treatment using
the thermal energy equation in the following form
\begin {equation}
\label{eq:energy}
 \doverdt{\Sigma c\subscr{v} T} + \nabla \cdot (\Sigma c\subscr{v} T {\bf u} )
       =  - p  \nabla \cdot {\bf u}  +  D  - Q  - 2 H \nabla \cdot \vec{F}.
\end{equation}
Here ${\bf u} = (u_r, u_{\varphi})$ is the two-dimensional velocity,
$\Sigma$ the density, $p$ the pressure, $T$ the (mid-plane)
temperature of the disk, and $c\subscr{v}$ is the specific heat
at constant volume. On the right hand side, the first term describes
compressional heating, $D$ the (vertically averaged) dissipation function,
$Q$ the local radiative cooling from the (vertical) surfaces of the disk, and
$\vec{F}$ denotes the two-dimensional radiative flux in the $(r,
\varphi)$-plane.  For all models $H$ is calculated from the
sound-speed as $H(r) = c\subscr{s} \, / \,
\Omega\subscr{K}(r)$, where $\Omega\subscr{K}$ is the Keplerian
angular velocity, and $c\subscr{s}=\sqrt{\gamma p / \Sigma}$,
$\gamma =1.43$ being the adiabatic index. The 2D-pressure is given by
$p= R_{gas} \Sigma T /\mu$ with the mean molecular weight $\mu=2.35$.

To calculate the radiative
losses $Q$ (from the two sides of the disk) we follow
\citet{2003ApJ...599..548D} and \citet{2005A&A...437..727K},
\[
    Q  =  2  \sigma\subscr{R}  T\subscr{eff}^4 \ ,
\nonumber
\]
where $\sigma\subscr{R}$ is the Stefan-Boltzmann constant and
$T\subscr{eff}$ is an estimate for the effective temperature
\citep{1990ApJ...351..632H}, given by\,:
\[
    T^4\subscr{eff} \, \tau\subscr{eff} = T^4 \ ,  \qquad  \mbox{with}
 \qquad   \tau\subscr{eff}
   = \frac{3}{8} \tau  + \frac{\sqrt{3}}{4} + \frac{1}{4 \tau} \ .
\]
For our two-dimensional disk we approximate the mean vertical optical
depth by \, $\tau =  (1/2) \kappa  \Sigma$,
where for the Rosseland mean opacity $\kappa$ the analytical formulae
by \citet{1985prpl.conf..981L} are adopted.  The radiative transport
in the plane of the disk is treated in the flux-limited diffusion
approximation where the flux is given by\,:
\[
\label{eq:raddif}
 \qquad   \vec{F}  =  -  \frac{\lambda c \, 4 a T^3}{\rho \kappa} \, \nabla T.
\]
Here $c$ is the speed of light, $a$ the radiation constant, $\rho =
\Sigma / (2 H)$ the mid-plane density, and $\lambda$ the flux-limiter
\citep[see][]{1989A&A...208...98K}.

In the following we present results of numerical simulations using
various sub-parts of the energy equation, Eq.~(\ref{eq:energy}). If
only the first term on the right hand side is used, the model is
called {\it adiabatic}\,; when only the last term is omitted, it is a
model with {\it local heating and cooling}\,; and the usage of the
full energy equation is termed {\it fully radiative}.  To make contact
with previous results we will also use {\it isothermal} models where a
pre-described radial temperature distribution is held fixed and no
energy equation is solved at all.  Please note that in the full
version, the radiative treatment (in Eq.~(\ref{eq:energy})\,)
incorporates full 3D effects of the radiative transport in that the
vertical part ($z$-direction) is taken care of by the local cooling
term, $Q$, and the horizontal part through $\vec{F}$.

\subsection{Reference Model}
The two-dimensional ($r - \varphi$) computational domain consists of a
complete ring of the protoplanetary disk centred on the star, extending
from $r\subscr{min} = 0.4$ to $r\subscr{max} = 2.5$ in units of $r_0 = a_{Jup} = 5.2$AU.
The mass of the central star is one solar mass, and the total disk mass inside 
$[r\subscr{min},r\subscr{max}]$ is $M_{disk} = 0.01 M_\odot$. 
For the present study we have kept the
kinematic viscosity constant with $\nu = 10^{15}$cm$^2$/s, a value which
relates to an equivalent $\alpha$ at $r_0$ of 0.004 for a disk aspect ratio of
$H/r = 0.05$, where $\nu = \alpha H^2 \Omega_K$.

To construct a {\it joint reference} model for the different types of approximations
to the energy equation (isothermal, adiabatic, etc.) we construct a {\it fully
radiative} model with no embedded planet, using the above layout.

\begin{figure}[ht]
\def\capfrac{1} 
\resizebox{0.49\linewidth}{!}{%
\includegraphics{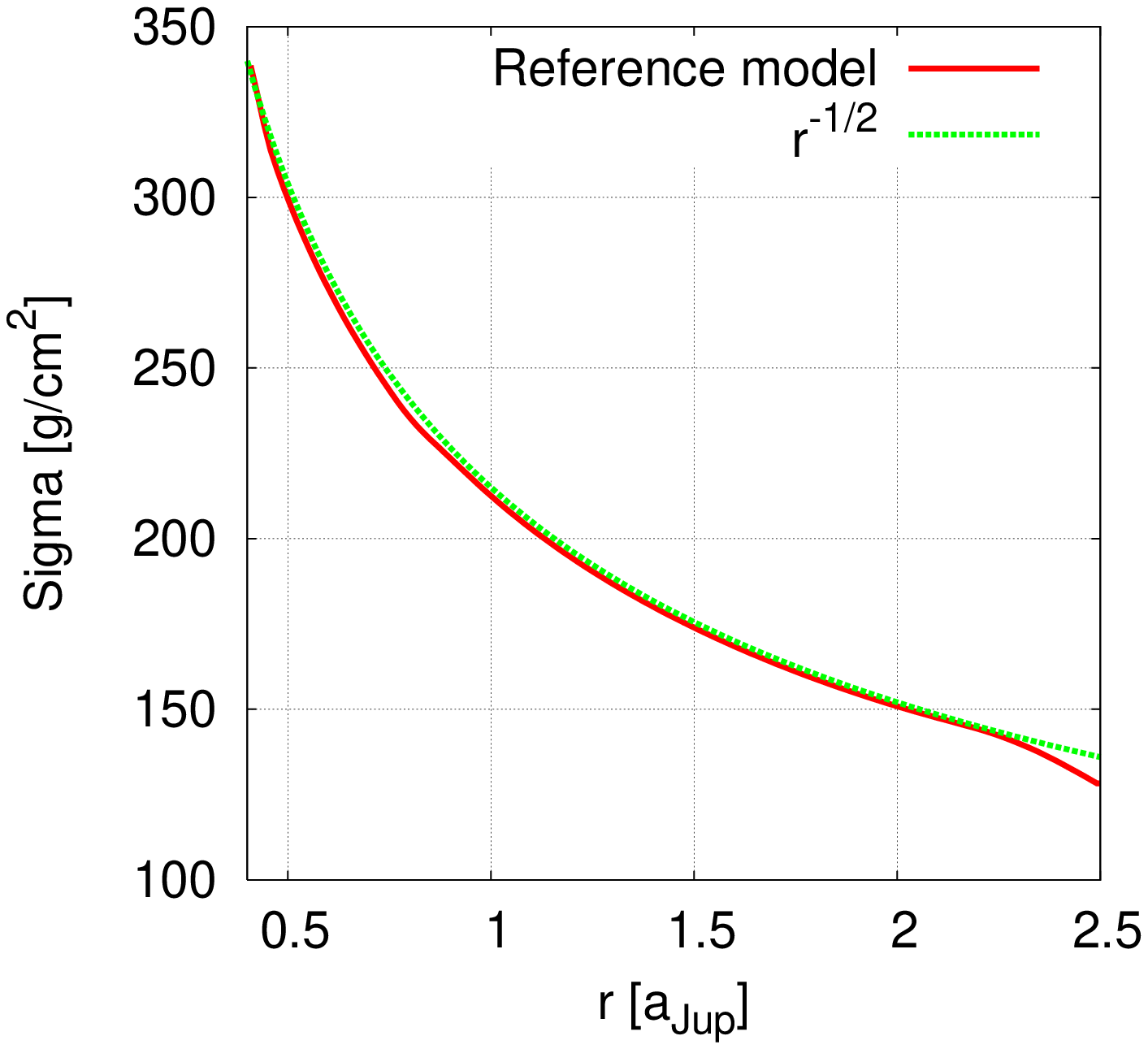}}
\resizebox{0.49\linewidth}{!}{%
\includegraphics{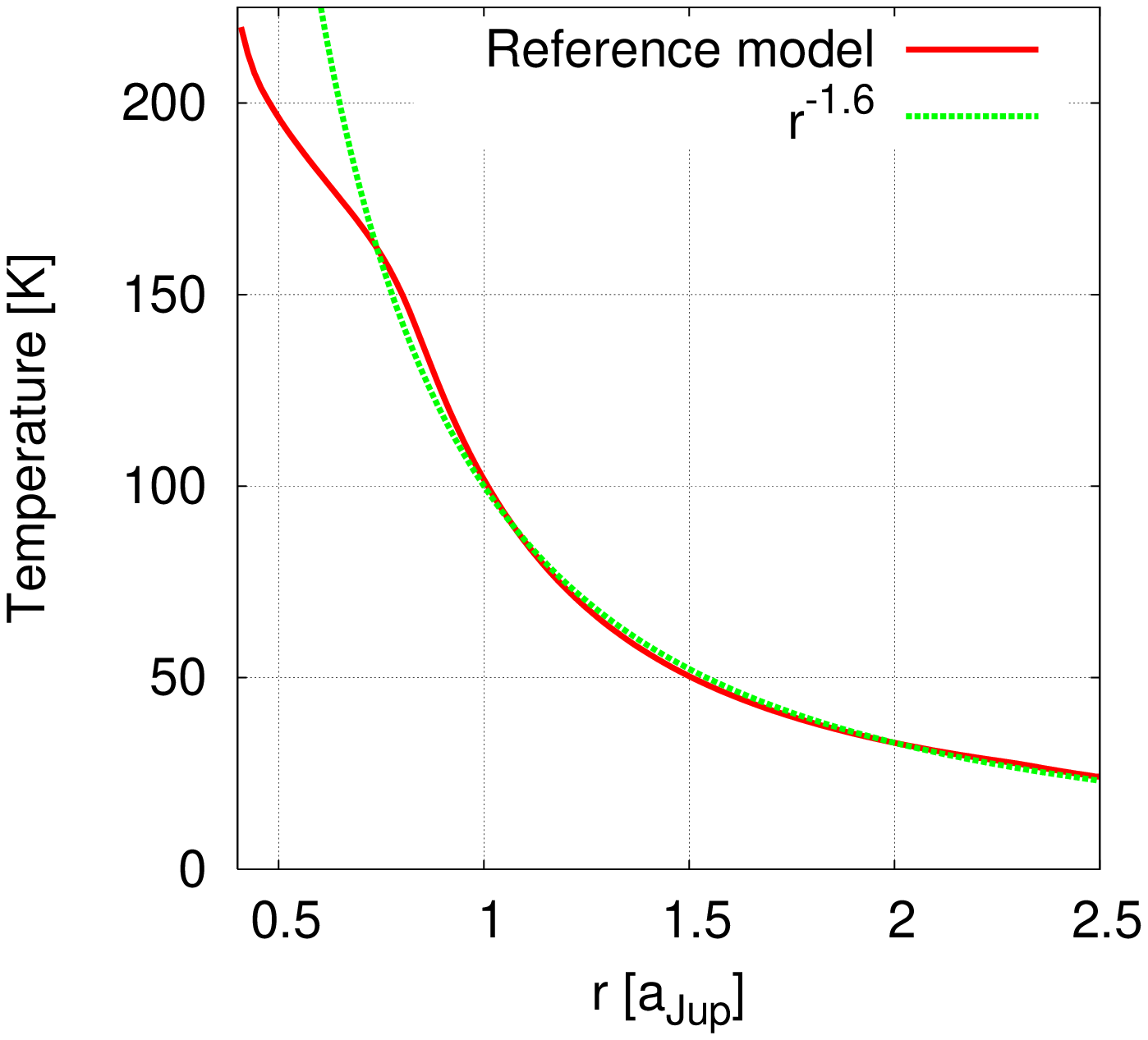}} 
  \caption{
Density and temperature vs. radius for the relaxed reference model.
For illustrative purpose, simple comparison power laws have been added. 
  }
   \label{fig:a02-sigtemp}
\end{figure}

This model is obtained from an approximate initial state (with $H/r=0.05$)
that is then relaxed
to its equilibrium by performing a time evolution of the disk solving
the full Navier-Stokes equations including the energy.
The initial disk stratification at the start of this process is given by
$\Sigma(r) \propto \, r^{-1/2}$,
$T(r) \propto r^{-1}$, and pure Keplerian rotation 
($u_r=0, u_\varphi = (G M_*/r)^{1/2}$). The evolution towards the
equilibrium takes about 150 orbits and the resulting density and temperature
distribution is displayed in Fig.~\ref{fig:a02-sigtemp}.
The density has a $\Sigma \propto r^{-1/2}$ profile which follows for constant $\nu$
directly from the angular momentum equation.
The temperature has a $T  \propto r^{-1.6}$ profile which follows 
from $Q = D$ for the used opacity. 
These relations are superimposed in Fig.~\ref{fig:a02-sigtemp}.
For small radii (higher temperatures) the opacity law has different temperature
dependence and the slope $T(r)$ changes.
The relative scale height for this radiative model is not constant but falls off
with radius. At $r=1$ we find $H/r \approx 0.045$ and at the outer radius
$H/r \approx 0.03$.

\subsection{Numerics}
We work in a rotating reference system, rotating with the orbital frequency
of the planet.
For our standard cases we use an equidistant grid in $r$ and $\varphi$ with
a resolution of $128 \times 384$.
In an effort to  ensure a  uniform environment for all models and
minimize disturbances (wave reflections) from the radial boundaries
we impose at $r\subscr{min}$ and $r\subscr{max}$ damping boundary
conditions where both velocity components are relaxed towards their initial
state on a timescale of the order of the local orbital period.
As the initial radial velocity is vanishing, this damping routine
ensures that no mass flows through the boundaries such that the total disk
mass remains constant. The angular velocity is relaxed towards the
Keplerian values.
For the density and temperature we apply closed radial boundary conditions.
In the azimuthal direction, periodic boundary conditions are imposed for all
variables.

The numerical details of the used finite volume code ({\tt RH2D})
relevant for these disk simulations have been described in
\citet{1999MNRAS.303..696K}, where we have additionally implemented
the {\tt FARGO} method \citep{2000A&AS..141..165M} that speeds up
the numerical computation of differentially rotating flows.  The
energy equation Eq.~(\ref{eq:energy}) is solved explicit-implicitly
applying operator-splitting.  The heating and cooling term $D-Q$ is
treated as one sub-step in this procedure
\citep[see][]{2004A&A...423..559G}, and the additional radiative
diffusion part in the energy equation is solved through an implicit
method to avoid possible time step limitations.

\begin{figure}
 \centering
 \includegraphics[width=0.9\linewidth]{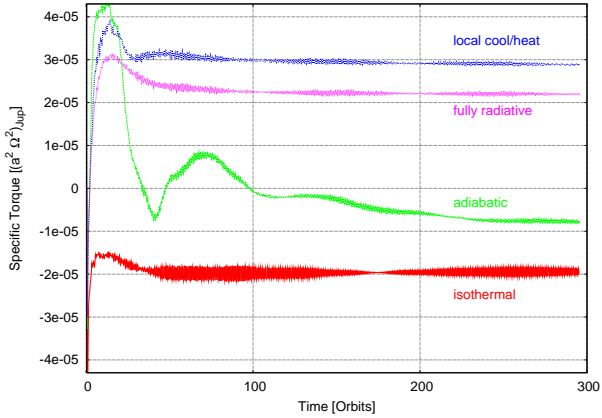}
 \caption{Time evolution of the specific total torque
  exerted by the disk on an
  embedded protoplanet of 20$M_{\rm Earth}$ for various approximations to the
  energy equation.
   \label{fig:r20b03-t0c}
   }
\end{figure}

\section{Results}
\label{sec:results}
\subsection{A model with 20~$M_{\rm Earth}$}
To illustrate the influence of the 4 different formulations of the energy
equation on the torque, we present a model with an embedded 
20~$M_{\rm Earth}$ planet. For this particular mass, a recent 3D-study to analyze
the effects of an isothermal disk on all orbital elements, indicated
negative rates for migration as well as for eccentricity and
inclination changes \citep{2007A&A...473..329C}. The starting
configuration (at $t=0$) for the 4 cases has been the equilibrated
reference model corresponding to the fully radiative case as described
above (see Fig.~\ref{fig:a02-sigtemp}). In Fig.~\ref{fig:r20b03-t0c}
we display the time evolution of the torque acting on the planet, for
the different formulations of the energy equation.  As expected, the
isothermal condition leads after about 50 orbital periods to a
constant negative torque implying inward migration.  The adiabatic
model has an initial phase of positive torques, before settling to
negative values of about 40\% of the isothermal case. This behaviour
has been suggested by \citet{2008ApJ...672.1054B} who argue that in
the adiabatic case the torques will be unsaturated and positive 
during the onset phase, while long-term calculations should converge
to negative values.

In contrast, both radiative models settle to constant positive values.
Here, the fully radiative case yields a 25\% smaller value due to the
included heat diffusion in the disk plane. In all cases, we have
continued the models with double and quadruple resolution. In general,
we find (for this planet mass) similar results and a convergence of
the torques at about double resolution. Only the model with pure
heating/cooling (i.e. no radiative diffusion) apparently requires much
higher resolution for convergence, a fact we attribute to the highly
local character of this type of energy equation. The addition of
radiative diffusion in the disk plane eases the numerical requirements
and makes at the same time the simulations physically more realistic.
\begin{figure}
\def\capfrac{1}
\resizebox{0.90\linewidth}{!}{%
\includegraphics[angle=270,width=0.7\linewidth]{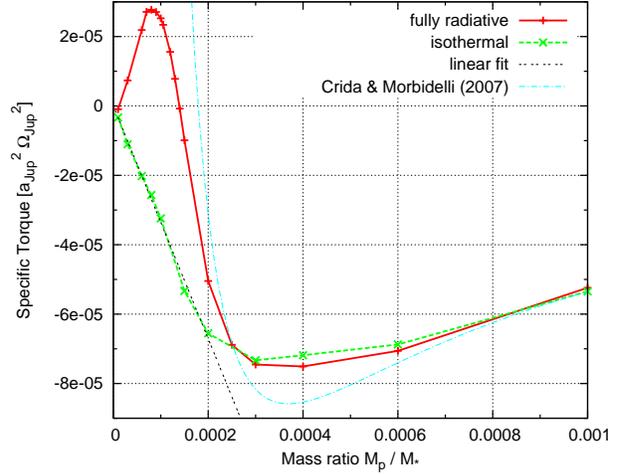}}
\\ \ 
\\ \ 
\resizebox{0.90\linewidth}{!}{%
\includegraphics[angle=270,width=0.7\linewidth]{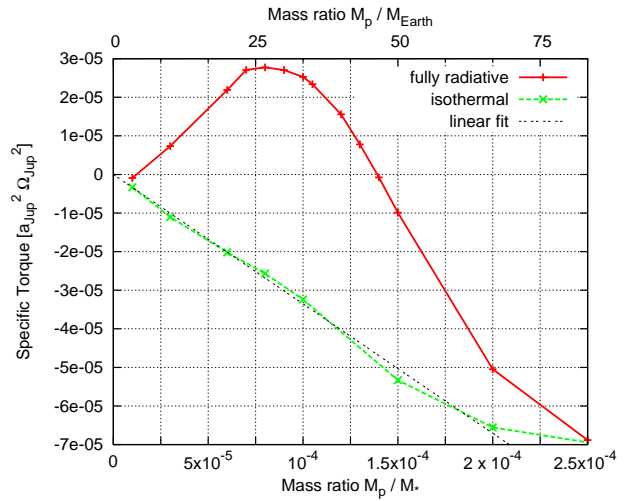}}
  \caption{Comparison of the specific torque exerted on an embedded
protoplanet for the isothermal and fully radiative model (thick
lines). The lower panel is an enlargement of the upper.  Two
analytic curves are superimposed (thin lines). }
   \label{fig:tcomp1}
\end{figure}

\subsection{Variation of the planet mass}
To estimate the influence of the planet mass we performed a sequence
of models with masses varying from $q=10^{-5}$ ($3 M_{\rm Earth}$) up
to $q=10^{-3}$ ($M_{\rm Jupiter}$). The resulting equilibrium 
specific torques are displayed in Fig.~\ref{fig:tcomp1} for the fully
radiative and isothermal case. Clearly, for small planetary masses up
to about $50 M_{\rm Earth}$ the torque exerted on the planet is
positive for the radiative model and turns negative above this mass
value (see bottom panel). On the contrary, for the isothermal models
the torques are negative throughout and follow the expected well known
results. From the superimposed dotted line, it is clear that the
specific torque is proportional to $q$, as predicted by linear theory
\citep[eg.][]{2002ApJ...565.1257T}.

For large planets ($q\geqslant 2.5\times 10^{-4}$), there is little
difference between radiative and isothermal disks, and the torque is
well approximated by the model given by \citet{2007MNRAS.377.1324C}\,:
the dot-dashed curve in the top panel comes from their Eq.~(15), with
$v_r$ replaced by $\frac{3}{2}\frac{\nu}{r}$. This model refines
the type~II migration rate for planets whose gap is not completely
empty\,; therefore, it applies here for $q \gtrsim 3\times 10^{-4}$
(see the gap profiles in Fig.~\ref{fig:sig-q}).  We find that this
model, and type~II migration in general, is also valid in radiative
disks.

\begin{figure}[ht]
\def\capfrac{1} 
\resizebox{0.49\linewidth}{!}{%
\includegraphics{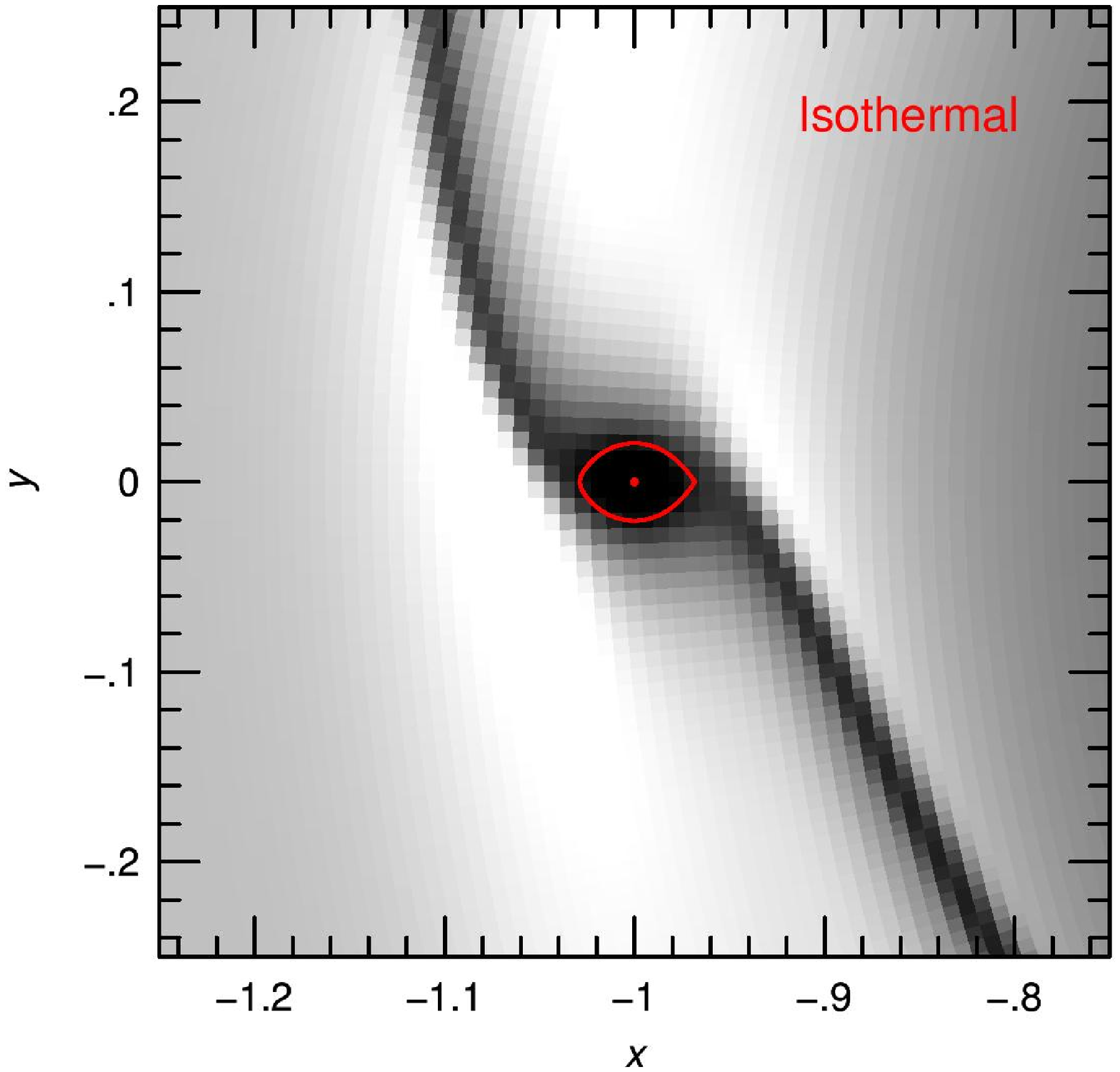}}
\resizebox{0.49\linewidth}{!}{%
\includegraphics{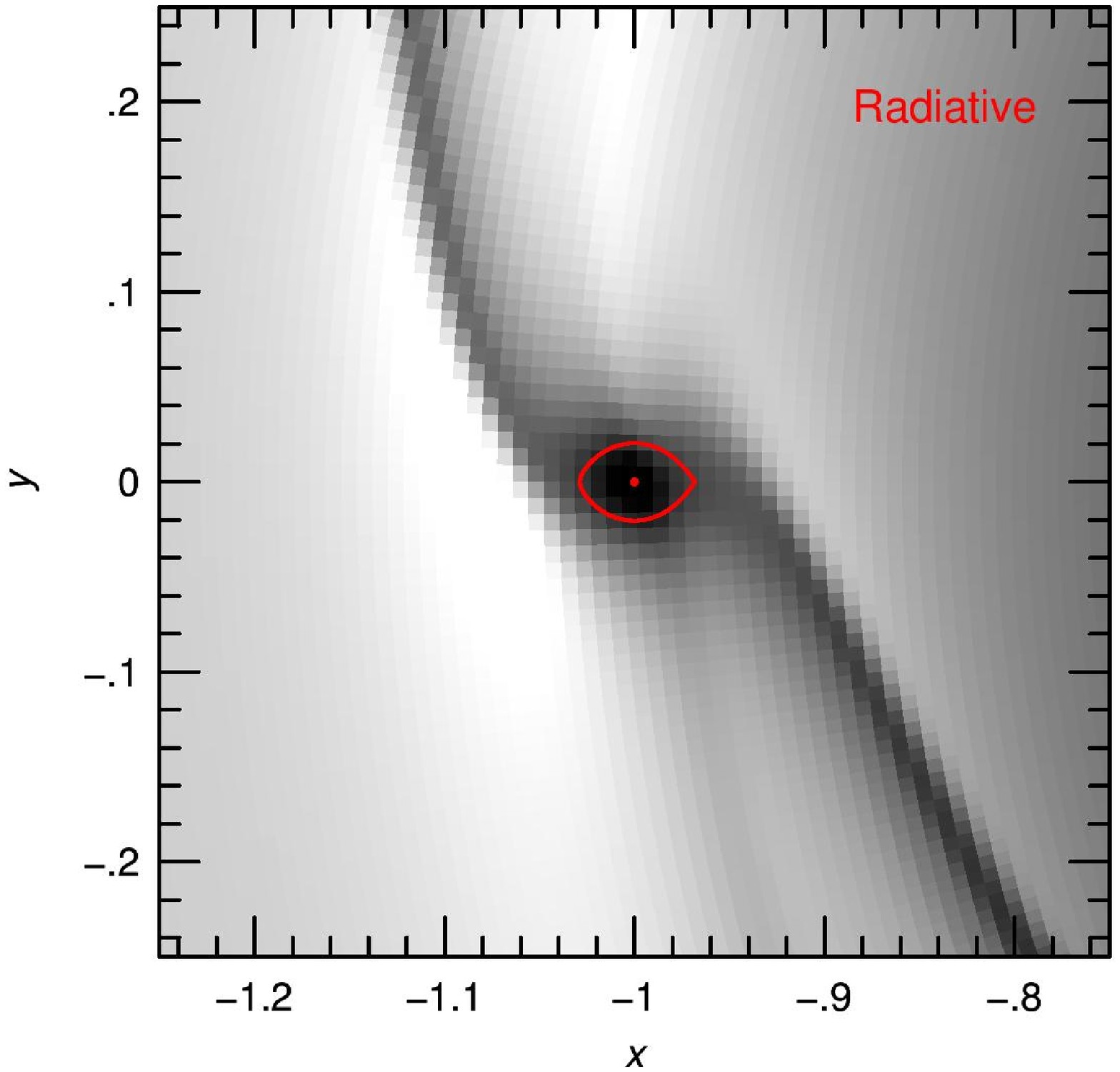}}
  \caption{ Two-dimensional grey-scale plot of the density ($\propto
\Sigma^{1/4}$-scaling between 150 and 350 g/cm$^{2}$) in the vicinity
of the planet for the isothermal model ({\bf left}) and the fully
radiative model ({\bf right}). The planet mass in this case refers to
$10^{-4} M_\odot$ or $33 M_{\rm Earth}$.  For these plots we have used
double resolution models ($256\times 768$).  }
   \label{fig:sig2d}
\end{figure}

The two-dimensional density distribution for the two cases (isothermal and fully radiative)
is displayed in Fig.~\ref{fig:sig2d} for a planetary mass of $q=10^{-4}$ ($33 M_{Earth}$)
which has the maximum positive torque. In the radiative model the higher temperature
results in slightly larger opening angles of the spiral arms with a reduced density, while
the isothermal situation allows for a higher mass concentration within the Roche-lobe
of the planet.
As expected by previous investigations
\citep{2008ApJ...672.1054B, 2008arXiv0804.4547P}, the inner half of the horseshoe region
ahead of the planet (in Fig.~\ref{fig:sig2d} below the planet)
is denser in the radiative case than in the
isothermal one, due to the incoming of colder gas from the outer half
after an U-turn. Symmetrically, the outer half of the horseshoe region
is slightly depleted behind the planet.  As a consequence of these
changes in density, the Lindblad torques are slightly reduced and the
corotation torque is more positive. As the net result we find for this
case an outward migration. 
To sustain the necessary unsaturated corotation torque on the long run, the 
libration timescales have to be comparable to the cooling times
\citep{2008ApJ...672.1054B, 2008arXiv0804.4547P}.
For our parameters we find a cooling time
$\tau_{cool} = T \Sigma c_v/Q$ of about 200 $\Omega_K^{\ -1}$ at the location $r_p$ of the planet.
This is comparable to the horseshoe libration time, $\tau_{lib} = 8 \pi r_p / (3 \Omega_K x_s)$
with the half-width of horseshoe-orbit,
$x_s = 1.16 r_p \sqrt{ q /(H/r)}$ \citep{2008ApJ...672.1054B}. In our case, for
$q=10^{-4}$ we find $\tau_{lib} = 150 \Omega_K^{\ -1}$.

\begin{figure}[ht]
\def\capfrac{1} 
\resizebox{0.49\linewidth}{!}{%
\includegraphics{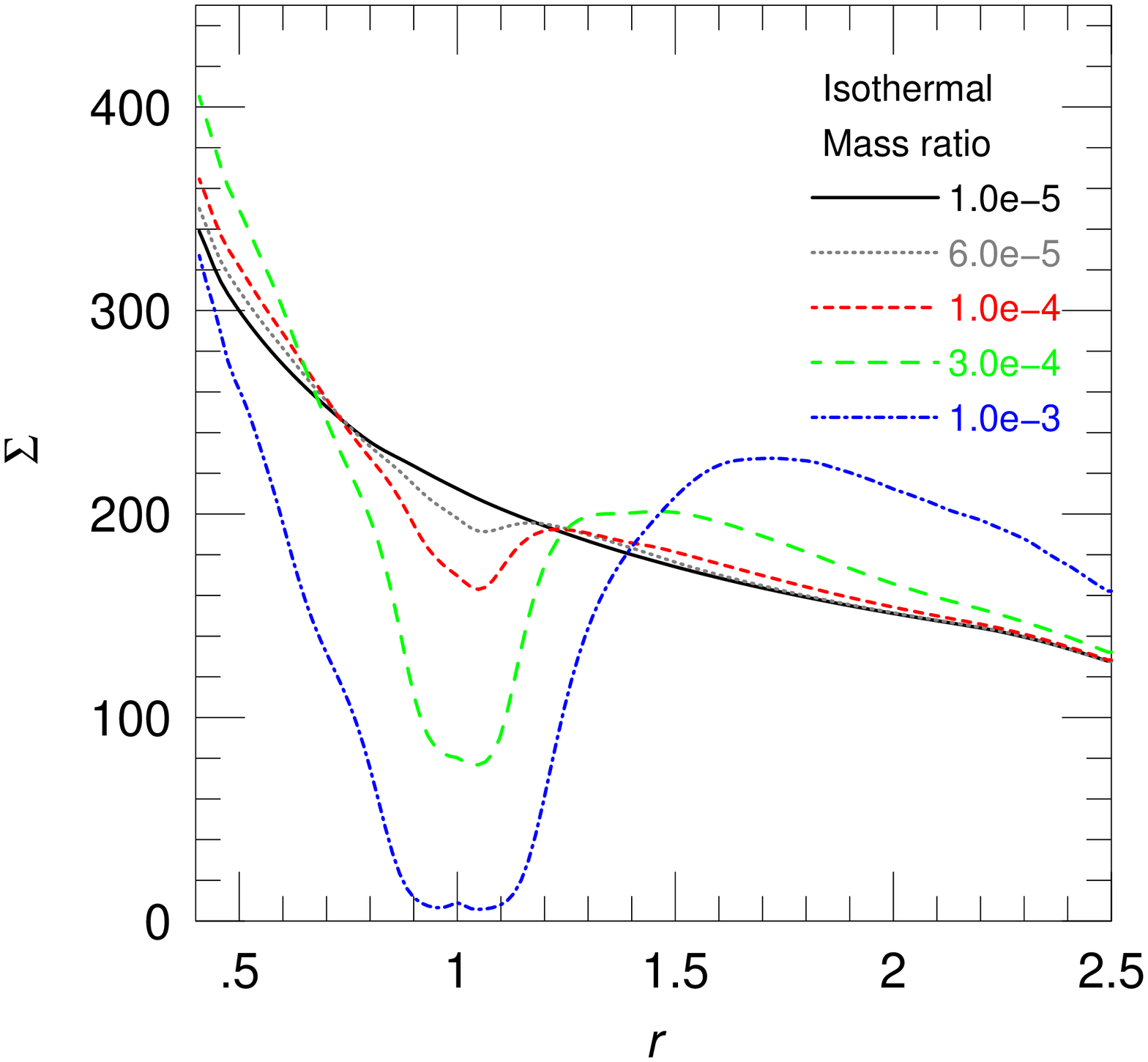}}
\resizebox{0.49\linewidth}{!}{%
\includegraphics{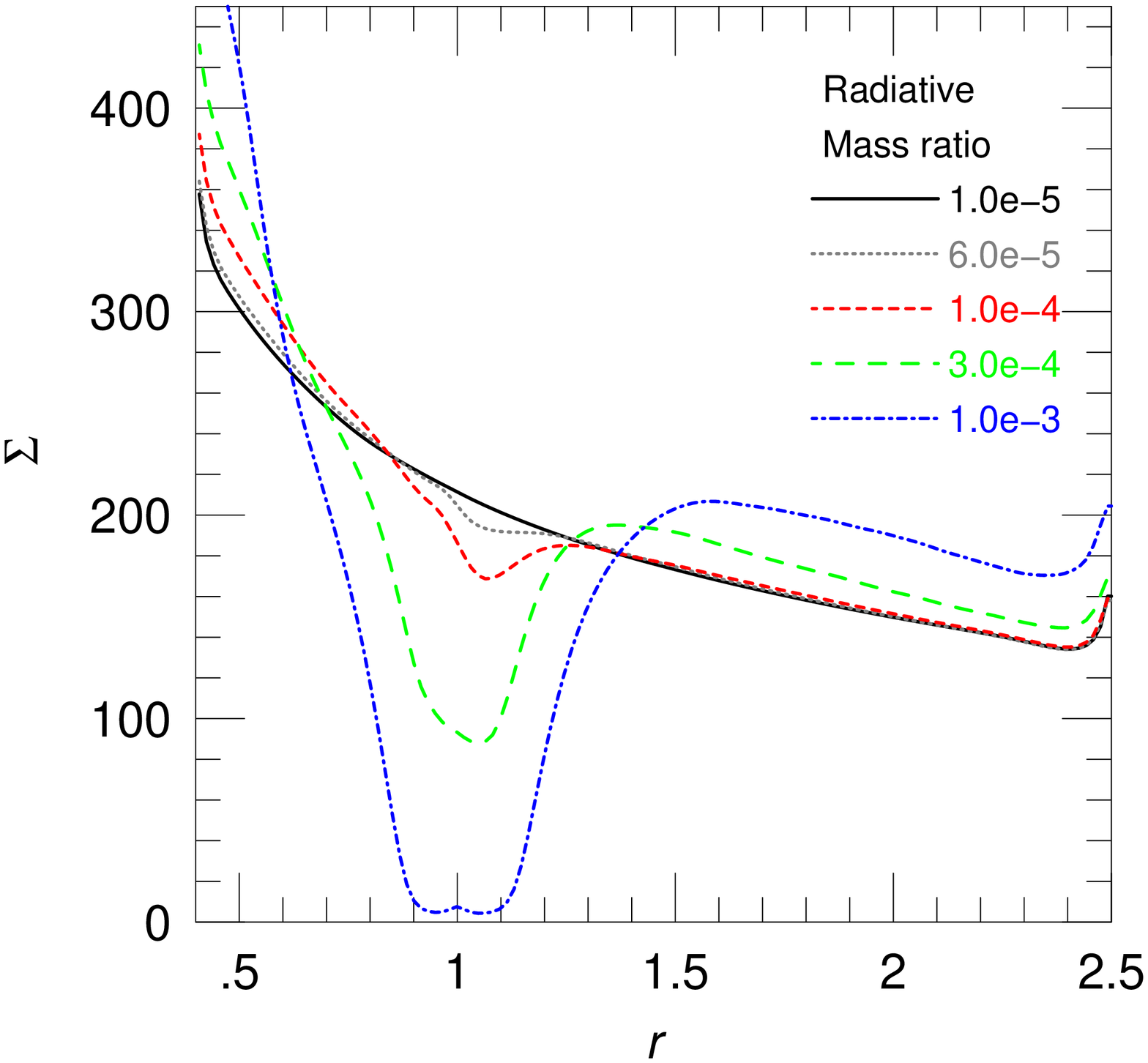}}
  \caption{
Azimuthally averaged density for isothermal and radiative cases for different planet masses.
  }
   \label{fig:sig-q}
\end{figure}
A comparison of the radial density stratification of the isothermal and radiative
cases (Fig.~\ref{fig:sig-q}) indicates shallower gaps for the latter, in
particular the region inside of the planet ($r < 1$) is less cleared. Larger planet
masses lead to gap opening with similar gap depths for the isothermal and radiative case.
For the largest planet mass the gap seems to be wider in the isothermal case, possibly due to
lower temperatures, but boundary effects (at $r_{min}$ and $r_{max}$) begin to become visible.
For larger planet masses a radially increased domain is clearly required. 
\section{Summary}
\label{sec:summary}
We have performed an investigation of the migration of planets in
disks using two-dimensional numerical simulation including
heating/cooling effects as well as radiative diffusion.

Using different formulations of the energy equation, we first show
that for a planet mass of $20 M_{\rm Earth}$, migration is directed
inwards in the isothermal and adiabatic situation, while inclusion of
radiative effects leads to an outward migration of the planet. This
finding supports the torque reversal mechanism in radiative disks due
to corotation effects as suggested by \citet{2008ApJ...672.1054B}.  A
detailed parameter study for planetary masses in the range between
$10^{-5}$ and $10^{-3} M_\odot$ shows that the effect is limited to
planets in the low mass regime, $M_p \leq 50 M_{\rm Earth}$ where
corotation effects are indeed important.  Larger mass planets open up
gaps in the disk and the migration rate becomes similar to the
isothermal case.

Our findings are particularly important for the first growth phase of
planets and may ease the problem to too rapid inward type~I
migration. Depending on the mass accretion rate onto the planet the
growing planetary embryos can spend an extended time span in an
outward migration phase and avoid loss into the star.
However, close-in planets exist for a range of planetary masses, according to the observations.
Thus, a significant and long outward migration phase may create new difficulties. Whether this
problem really exists can only be answered by following the actual long-term migration of planets
through the disk including its mass growth.

Constructing the necessary migration histories of planetary cores, to
be used in population synthesis models, requires suitable scaling laws
for the migration process as a function of disk parameter
($\Sigma(r)$, $T(r)$) for realistic accretion disks with net mass
flow.  The present study can be used as a starting point for these
larger parameter studies. The inclusion of three-dimensional effects
and additional physics (MHD, self-gravity, mass accretion) will make
the models even more realistic in the future.
 
\begin{acknowledgements}

Very fruitful discussions with Fred\'eric Masset and Cl\'ement Baruteau
are gratefully acknowledged.
A. Crida acknowledges the support through the German Research
Foundation (DFG) grant KL 650/7.

\end{acknowledgements}

\bibliographystyle{aa}
\bibliography{kley8}

\begin{thebibliography}{26}
\expandafter\ifx\csname natexlab\endcsname\relax\def\natexlab#1{#1}\fi

\bibitem[{{Alibert} {et~al.}(2004){Alibert}, {Mordasini}, \&
  {Benz}}]{2004A&A...417L..25A}
{Alibert}, Y., {Mordasini}, C., \& {Benz}, W. 2004, \aap, 417, L25

\bibitem[{{Baruteau} \& {Masset}(2008)}]{2008ApJ...672.1054B}
{Baruteau}, C. \& {Masset}, F. 2008, \apj, 672, 1054

\bibitem[{{Cresswell} {et~al.}(2007){Cresswell}, {Dirksen}, {Kley}, \&
  {Nelson}}]{2007A&A...473..329C}
{Cresswell}, P., {Dirksen}, G., {Kley}, W., \& {Nelson}, R.~P. 2007, \aap, 473,
  329

\bibitem[{{Crida} \& {Morbidelli}(2007)}]{2007MNRAS.377.1324C}
{Crida}, A. \& {Morbidelli}, A. 2007, \mnras, 377, 1324

\bibitem[{{Crida} {et~al.}(2008){Crida}, {S{\'a}ndor}, \&
  {Kley}}]{2008A&A...483..325C}
{Crida}, A., {S{\'a}ndor}, Z., \& {Kley}, W. 2008, \aap, 483, 325

\bibitem[{{D'Angelo} {et~al.}(2003){D'Angelo}, {Henning}, \&
  {Kley}}]{2003ApJ...599..548D}
{D'Angelo}, G., {Henning}, T., \& {Kley}, W. 2003, \apj, 599, 548

\bibitem[{{Goldreich} \& {Tremaine}(1979)}]{1979ApJ...233..857G}
{Goldreich}, P. \& {Tremaine}, S. 1979, \apj, 233, 857

\bibitem[{{G{\"u}nther} {et~al.}(2004){G{\"u}nther}, {Sch{\"a}fer}, \&
  {Kley}}]{2004A&A...423..559G}
{G{\"u}nther}, R., {Sch{\"a}fer}, C., \& {Kley}, W. 2004, \aap, 423, 559

\bibitem[{{Hubeny}(1990)}]{1990ApJ...351..632H}
{Hubeny}, I. 1990, \apj, 351, 632

\bibitem[{{Ida} \& {Lin}(2008)}]{2008ApJ...673..487I}
{Ida}, S. \& {Lin}, D.~N.~C. 2008, \apj, 673, 487

\bibitem[{{Klahr} \& {Kley}(2006)}]{2006A&A...445..747K}
{Klahr}, H. \& {Kley}, W. 2006, \aap, 445, 747

\bibitem[{{Kley}(1989)}]{1989A&A...208...98K}
{Kley}, W. 1989, \aap, 208, 98

\bibitem[{{Kley}(1999)}]{1999MNRAS.303..696K}
---. 1999, \mnras, 303, 696

\bibitem[{{Kley} {et~al.}(2005){Kley}, {Lee}, {Murray}, \&
  {Peale}}]{2005A&A...437..727K}
{Kley}, W., {Lee}, M.~H., {Murray}, N., \& {Peale}, S.~J. 2005, \aap, 437, 727

\bibitem[{{Korycansky} \& {Pollack}(1993)}]{1993Icar..102..150K}
{Korycansky}, D.~G. \& {Pollack}, J.~B. 1993, Icarus, 102, 150

\bibitem[{{Lin} \& {Papaloizou}(1985)}]{1985prpl.conf..981L}
{Lin}, D.~N.~C. \& {Papaloizou}, J.~C.~B. 1985, in Protostars and Planets II,
  981

\bibitem[{{Masset}(2000)}]{2000A&AS..141..165M}
{Masset}, F. 2000, \aaps, 141, 165

\bibitem[{{Masset} {et~al.}(2006{\natexlab{a}}){Masset}, {D'Angelo}, \&
  {Kley}}]{2006ApJ...652..730M}
{Masset}, F.~S., {D'Angelo}, G., \& {Kley}, W. 2006{\natexlab{a}}, \apj, 652,
  730

\bibitem[{{Masset} {et~al.}(2006{\natexlab{b}}){Masset}, {Morbidelli}, {Crida},
  \& {Ferreira}}]{2006ApJ...642..478M}
{Masset}, F.~S., {Morbidelli}, A., {Crida}, A., \& {Ferreira}, J.
  2006{\natexlab{b}}, \apj, 642, 478

\bibitem[{{Nelson}(2005)}]{2005A&A...443.1067N}
{Nelson}, R.~P. 2005, \aap, 443, 1067

\bibitem[{{Paardekooper} \& {Mellema}(2006)}]{2006A&A...459L..17P}
{Paardekooper}, S.-J. \& {Mellema}, G. 2006, \aap, 459, L17

\bibitem[{{Paardekooper} \& {Mellema}(2008)}]{2008A&A...478..245P}
---. 2008, \aap, 478, 245

\bibitem[{{Paardekooper} \& {Papaloizou}(2008)}]{2008arXiv0804.4547P}
{Paardekooper}, S.-J. \& {Papaloizou}, J.~C.~B. 2008, ArXiv:astro-ph/0804.4547,
  804

\bibitem[{{Tanaka} {et~al.}(2002){Tanaka}, {Takeuchi}, \&
  {Ward}}]{2002ApJ...565.1257T}
{Tanaka}, H., {Takeuchi}, T., \& {Ward}, W.~R. 2002, \apj, 565, 1257

\bibitem[{{Tanaka} \& {Ward}(2004)}]{2004ApJ...602..388T}
{Tanaka}, H. \& {Ward}, W.~R. 2004, \apj, 602, 388

\bibitem[{{Ward}(1997)}]{1997Icar..126..261W}
{Ward}, W.~R. 1997, Icarus, 126, 261

\end{thebibliography}
\end{document}